\pdfoutput=1

\documentclass[twocolumn,groupedaddress,superscriptaddress,10pt]{revtex4-1}

\usepackage[utf8]{inputenc}
\usepackage{enumerate}
\usepackage{amsmath}
\usepackage{braket}
\usepackage[margin=1in]{geometry}
\usepackage{graphicx}
\usepackage{siunitx}
\usepackage{upgreek}
\usepackage{xcolor}
\usepackage{hyperref}
\usepackage[version=4]{mhchem}
\DeclareUnicodeCharacter{2009}{ }

\begin{document}


\title{Formation of a lateral p-n junction light-emitting diode on an n-type high-mobility GaAs/Al$_{0.33}$Ga$_{0.67}$As heterostructure}

\author{C. P. Dobney} \thanks{Present address: University of Toronto Department of Physics, McLennan Physical Laboratories, 60 St. George Street, Toronto, ON M5S 1A7, Canada.\\ E-mail: pria.dobney@mail.utoronto.ca}
\affiliation{National Physical Laboratory, Hampton Road, Teddington TW11 0LW, United Kingdom}

\author{A. Nasir} 
\affiliation{National Physical Laboratory, Hampton Road, Teddington TW11 0LW, United Kingdom}
\author{P. See} 
\affiliation{National Physical Laboratory, Hampton Road, Teddington TW11 0LW, United Kingdom}
\author{C. J. B. Ford} 
\affiliation{Cavendish Laboratory, University of Cambridge, J. J. Thomson Avenue, Cambridge CB3 0HE, United Kingdom}
\author{J.~P.~Griffiths} 
\affiliation{Cavendish Laboratory, University of Cambridge, J. J. Thomson Avenue, Cambridge CB3 0HE, United Kingdom}
\author{C. Chen} 
\affiliation{Cavendish Laboratory, University of Cambridge, J. J. Thomson Avenue, Cambridge CB3 0HE, United Kingdom}
\author{D. A. Ritchie}
\affiliation{Cavendish Laboratory, University of Cambridge, J. J. Thomson Avenue, Cambridge CB3 0HE, United Kingdom}
\author{M. Kataoka}
\affiliation{National Physical Laboratory, Hampton Road, Teddington TW11 0LW, United Kingdom}

\date{\today}
\begin{abstract}
We have fabricated a device which includes two lateral p-n junctions on an n-type GaAs/Al$_{0.33}$Ga$_{0.67}$As heterostructure. A section of the n-type material has been converted to p-type by removing dopants and applying a voltage to a gate placed in this region. Controlled electroluminescence from both of the p-n junctions has been demonstrated by varying the applied bias voltages.
An emission peak with a width of $\sim 8$~nm is observed around 812 nm. The electroluminescence seen from both junctions is considered to originate from the GaAs quantum well layer in the device.
The lithographic techniques that we have developed are compatible for further integration of gated quantum devices such as single-electron pumps to build on-demand single-photon sources.
\end{abstract}

\maketitle


Recently, the possibility of forming a single-photon source from two-dimensional media and single electron transfer has been investigated \cite{foden2000high, hosey2004lateral, gell2006surface, hsiao2020single}. 
The idea is to inject electrons in a controlled manner across a p-n junction and therefore deterministically generate single-photon pulses on demand.
A lateral p-n junction can be formed of a region of two-dimensional electron gas adjacent to a region of two-dimensional hole gas. Single-photon emission occurs when, after being transported across the p-n junction, an electron recombines with a hole in the p-type region \cite{hsiao2020single}.

Several studies of different types of lateral p-n junction devices have been conducted in III-V semiconductor heterostructures, particularly GaAs/AlGaAs systems. In the focused ion molecular beam epitaxy method, two adjacent regions are selectively doped with Si and Be to create the n- and p-type regions \cite{vijendran1999independently}.
In the facet regrowth method, both p- and n-type regions are created using Si doped on different facets of the GaAs surface \cite{vaccaro1998light, north1999two}.
Cecchini \textit{et al.} formed a lateral p-n junction from a p-type substrate by etching away part of the Be-doped AlGaAs and forming an n-type AuGeNi contact. \cite{cecchini2003high, cecchini2004surface, cecchini2005surface}. Dai \textit{et al.} used two inducing gates to form the two-dimensional electron and hole gases \cite{dai2014high, dai2012lateral}.
Helgers \textit{et al.} used quantum wires on GaAs substrates as channels to transport optically excited electrons and holes using surface acoustic waves \cite{helgers2019sidewall}.
The formation of lateral p-n junctions is also possible in other types of material systems, such as two-dimensional materials \cite{frisenda2018atomically}.
The fabrication process of the lateral p-n junction can complicate further integration of additional components, such as single-electron sources, and may compromise the overall device performance. For example, single-photon emission has been demonstrated by Hsiao \textit{et al.} \cite{hsiao2020single} based on the inducing method \cite{dai2014high}, but quantised single-electron transfer was not observed in the same device (in principle, however, current quantisation should be possible \cite{chung2019quantized}).

In this letter, we present a lateral p-n junction device made from a standard n-type GaAs/AlGaAs wafer. We convert a selected part of the wafer to p-type by etching away the donor dopant layer and electrically inducing holes via a gate. Our method follows the idea developed by Mondal {\it et al.}\ to convert a part of doped n-type material into a dopant-free n-type region by field effect \cite{Mondal2014}. Our device features two lateral p-n junctions: one on the left-hand side of the device, one on the right-hand side.
We demonstrate controlled, on-demand electroluminescence from the p-n junctions and assign the emission to be from the GaAs quantum well by spectral studies. 
Starting from n-type GaAs, rather than undoped GaAs as is found in the literature, allows us to simplify the process of fabricating nano-device structures (such as single-electron pumps) on the n-type side by using existing methods, rather than having to devise new methods to do so \cite{buonacorsi2021non}.
Our method is advantageous in that it is compatible with standard nanodevice lithography, enabling easier integration of single-electron sources \cite{Blumenthal2007,Fletcher2013} and other device components.

\begin{figure*}
    \centering
    \includegraphics[width=\textwidth]{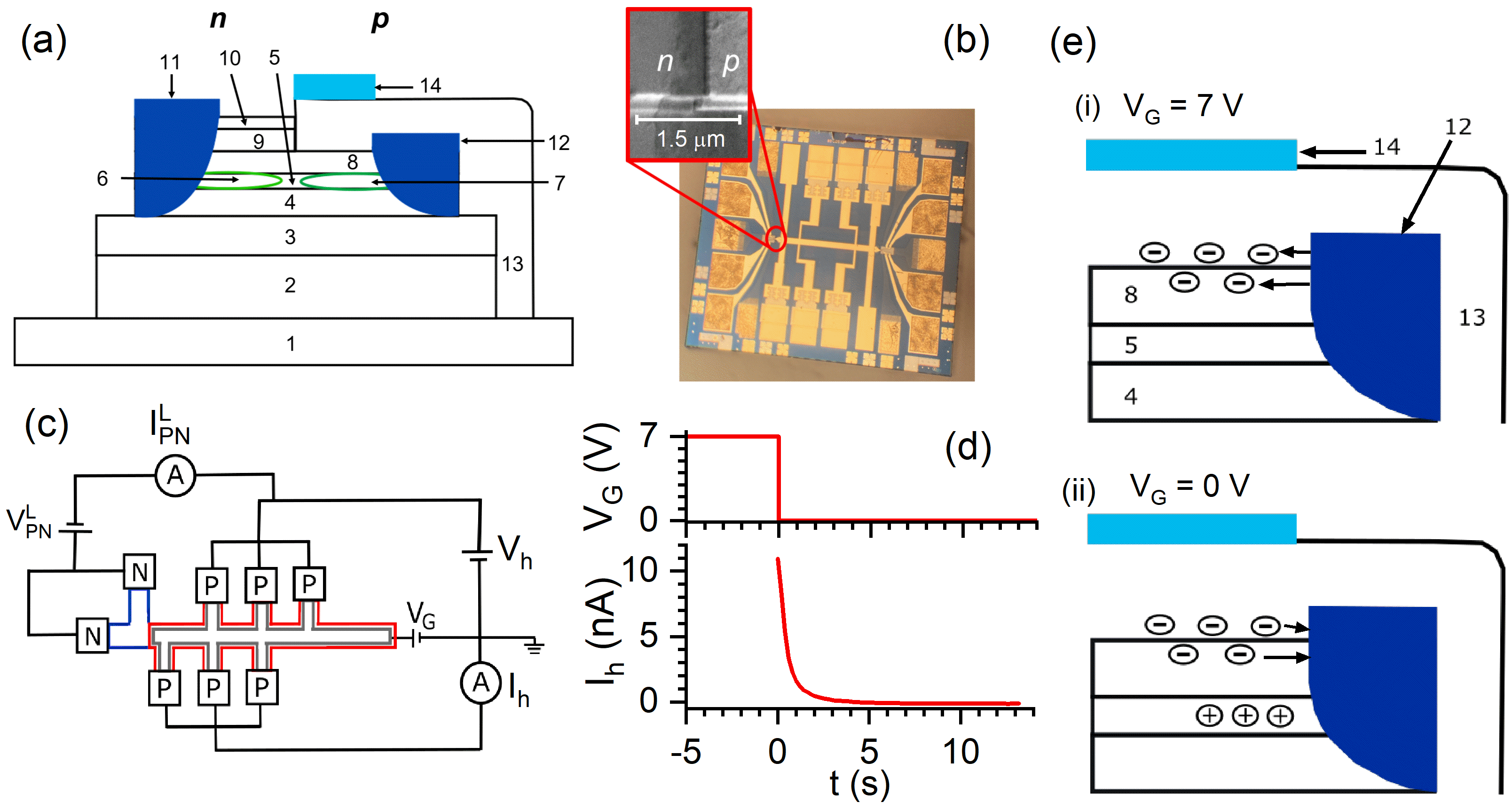}
    \caption{\scriptsize{
    \textbf{a:} Schematic cross-section of wafer showing device structure. 
    Layer 1: GaAs substrate; 2: 500~nm GaAs buffer; 3: 200~nm graded Al$_{0.05}$Ga$_{0.95}$As to Al$_{0.33}$Ga$_{0.67}$As layer; 4: lower 50~nm Al$_{0.33}$Ga$_{0.67}$As spacer; 5: 15~nm undoped GaAs quantum well; 6: two-dimensional electron gas; 7: induced two-dimensional hole gas; 8: upper 40~nm Al$_{0.33}$Ga$_{0.67}$As spacer; 9: 40~nm Si-doped Al$_{0.33}$Ga$_{0.67}$As; 10: 10 nm GaAs cap layer; 11: AuGeNi n-type contact; 12: AuBe p-type contact; 13: Al$_{2}$O$_{3}$ insulator; 14: Ti/Au inducing gate.
    \textbf{b:} Photograph of device; inset: scanning electron microscope image of the left-hand p-n junction region. 
    \textbf{c:} Circuit diagram (showing the left-hand p-n junction only) showing mesa layers for p-type (red outline) and n-type (blue outline) regions and the inducing gate (grey outline).
    \textbf{d:} Stepped gate voltage $V_{\rm{G}}$ and induced hole current $I_{\rm{h}}$.
    \textbf{e:} Suspected process of inducing build-up of charge in the device. \textbf{(i):} Applied gate voltage = 7~V, \textbf{(ii):} gate voltage stepped to 0~V.}
    }
    \label{fig:device}
\end{figure*}
The wafer is produced by molecular beam epitaxy, starting with a 500~nm GaAs buffer [Fig. 1(a) layer 2] and a 200~nm layer of graded Al$_{0.05}$Ga$_{0.95}$As to Al$_{0.33}$Ga$_{0.67}$As [Fig. 1(a) layer 3] on a semi-insulating GaAs substrate [Fig. 1(a) layer 1]. A 15~nm undoped GaAs quantum well [Fig 1(a) layer 5] is sandwiched between a lower 50~nm and upper 40~nm Al$_{0.33}$Ga$_{0.67}$As spacer [Fig. 1(a) layers 4 and 8, respectively]. The n-type carriers are provided by the 40~nm Si-doped Al$_{0.33}$Ga$_{0.67}$As [Fig. 1(a) layer 9], the only doped layer in the wafer, and the wafer is capped with 10~nm undoped GaAs [Fig. 1(a) layer 10]. The doping concentration for the 40~nm Si-doped Al$_{0.33}$Ga$_{0.67}$As layer [Fig. 1(a) layer 9] is \num{9.5e17} Si dopants per cm$^{3}$.
The electron carrier density is $n = \num{1.5e15}$~m$^{-2}$ and the electron mobility is $\mu = 100$~m$^{2}$/Vs, measured at 1.5~K.

\begin{figure}
    \centering
    \includegraphics[scale=0.39]{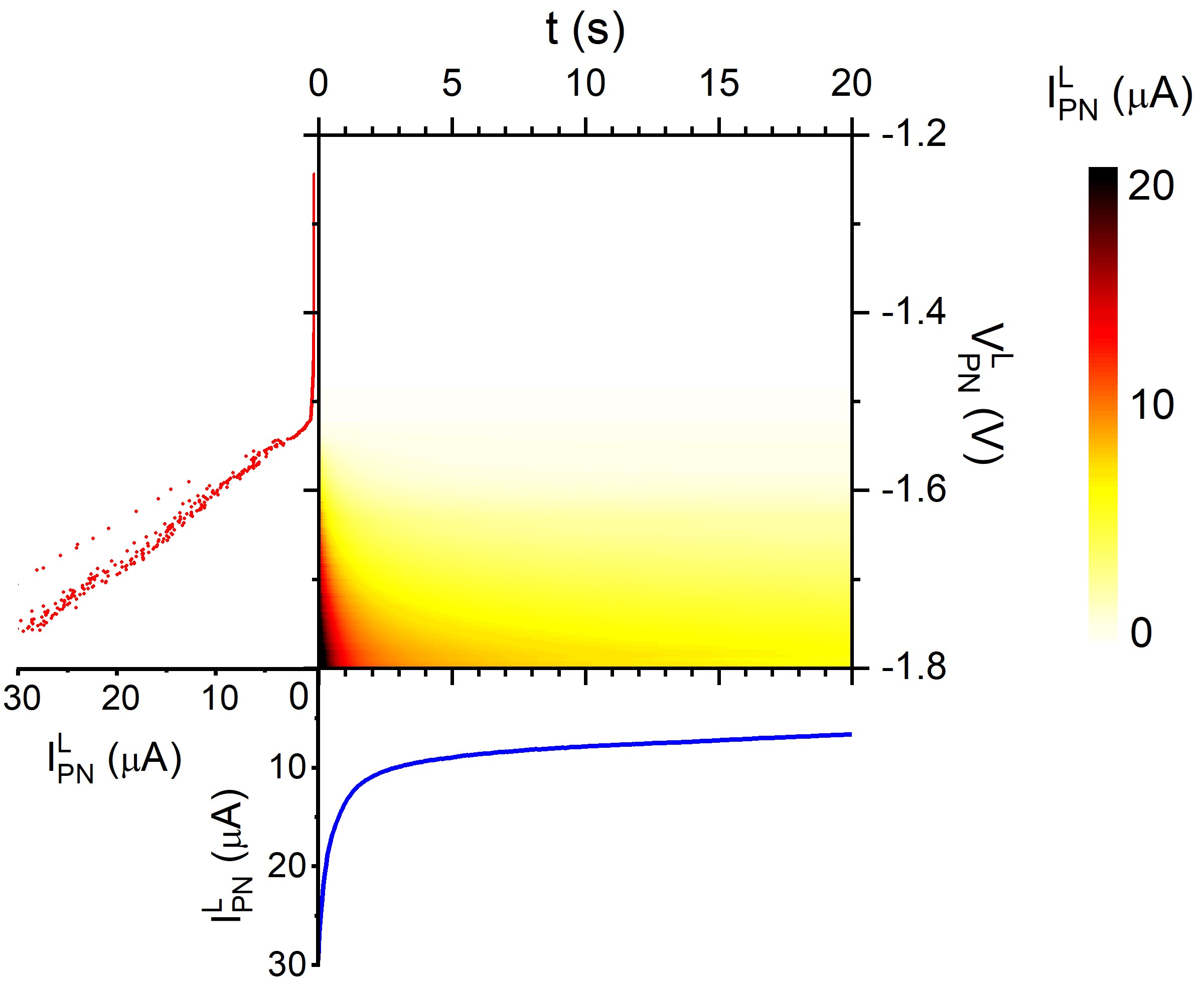}
    \caption{\scriptsize{{\textbf{Two-dimensional colour plot:} Time-dependent p-n junction current $I_{\rm{PN}}^{\rm{L}}$} as p-n junction bias $V_{\rm{PN}}^{\rm{L}}$ is varied from $-1.2$~V to $-1.8$~V. \textbf{Left panel:} Behaviour of $I_{\rm{PN}}^{\rm{L}}$ as $V_{\rm{PN}}^{\rm{L}}$ is varied, at $t = 0$~s. \textbf{Bottom panel:} Decay of $I_{\rm{PN}}^{\rm{L}}$ over time, at $V_{\rm{PN}}^{\rm{L}} = -1.8$~V. }
    }
    \label{fig:PNcurrent}
\end{figure}

Fabrication begins by removing the undoped 10~nm GaAs and 40~nm Si-doped Al$_{0.33}$Ga$_{0.67}$As layers using wet chemical etching to produce an intrinsic region in selected areas where we intend to form p-type regions. A typical GaAs/AlGaAs etchant was used - H$_{2}$SO$_{4}$:H$_{2}$O$_{2}$:H$_{2}$O in a concentration of 1:80:1600. 
Next follows a deeper isolation wet etch past the GaAs quantum well which is stopped partway in the lower Al$_{0.33}$Ga$_{0.67}$As spacer layer [Fig. 1(a) layer 4] in order to define a mesa to contain both types of carriers both in n- and p-type regions. n- and p-type contacts are then formed using AuGeNi [Fig. 1(a) layer 11] and AuBe [Fig. 1(a) layer 12] respectively by thermal evaporation and annealing, to contact two-dimensional electron and hole gases [Fig. 1(a) layers 6 and 7]. A 100 nm thick Al$_{2}$O$_{3}$ insulator is deposited by atomic layer deposition [Fig. 1(a) layer 13]. Windows in the Al$_{2}$O$_{3}$ insulator are then opened using buffered HF to expose the underlying Ti/Au bond pad contacts, which were formed by thermal evaporation after annealing the AuGeNi ohmic contacts and before depositing the Al$_{2}$O$_{3}$ insulator. Finally, Ti/Au inducing gates [Fig. 1(a) layer 14] are defined by electron beam lithography and aligned to the etched boundary with the n-type region within 100~nm accuracy.
Figure 1(b) shows a photograph of the device and the inset shows a scanning electron microscopy image of the left-hand p-n junction.
We use a bottom-loading dilution refrigerator (operating at 4 K) for characterising the lateral p-n junction. The probe is equipped with a fibre-coupled lens assembly and Attocube piezoelectric stages used for positioning and focusing. The device is mounted onto the probe, facing the lens. Voltages are supplied by Yokogawa GS-210s and currents are measured by Femto DDPCA-300s. The single-mode optical fibre from the probe is coupled to an Andor Shamrock 750~mm focal length Czerny-Turner spectrometer and electroluminescence spectra from the device are measured using an Andor Newton back-illuminated EMCCD camera. 
In order to form each p-n junction, a voltage $V_{\rm{G}}$ is applied to the gate [Fig. 1(a) layer 12] which induces a two-dimensional hole gas under the gate, adjacent to a two-dimensional electron gas. The formation of the hole gas is tested by applying a voltage $V_{\rm{h}}$ and measuring the current $I_{\rm{h}}$ induced between the pair of p-type ohmic contacts, as shown in Fig. 1(c). We found that in this particular device, if we apply a negative bias voltage onto the inducing gate, the hole gas can be induced only for a short time (less than 1 s), but the carrier density decays quickly.
We suspect that either the Al$_{2}$O$_{3}$ insulator layer or the interface between the insulator and the Al$_{0.33}$Ga$_{0.67}$As surface positively charges up, counteracting the gate voltage. We found that if we apply a positive bias to the gate prior to inducing the hole gas, then drop the voltage to zero, a hole gas is induced and can survive for around 1~s, long enough to take electroluminescence measurements (this behaviour has been reproduced by another device from the same fabrication batch). Figure 1(d) shows the hole current $I_{\rm{h}}$ as a function of time after $V_{\rm{G}}$ is stepped from 7~V to 0~V.
We believe that during the application of the positive bias, negative charges accumulate either in the insulator layer or at the insulator-Al$_{0.33}$Ga$_{0.67}$As interface [Fig. 1(e)(i)].
After $V_{\rm{G}}$ is set to zero, these negative charges slowly decay over a timescale of roughly 1~s [Fig. 1(e)(ii)]. While the negative charges are still present, the hole gas is induced in the GaAs quantum well. We have chosen this method to induce the hole gas for electroluminescence measurements.
The short lifetime of the induced hole gas makes the measurement of current-voltage characteristics difficult. However, it is possible to take an equivalent measurement by monitoring the time-dependent current as the p-n junction bias voltage is varied. Figure~\ref{fig:PNcurrent} shows the time-dependent p-n junction current $I_{\rm{PN}}^{\rm{L}}$ as a function of the p-n junction bias applied to the left-hand junction $V_{\rm{PN}}^{\rm{L}}$ [see Fig.~\ref{fig:device}(c)].
As $V_{\rm{PN}}^{\rm{L}}$ becomes more negative (forward bias) and passes the threshold of $-1.5$~V, the p-n junction current begins to flow. On the left panel in Fig.~\ref{fig:PNcurrent}, the value of the current at $t = 0$ is plotted. This current shows diode-like behaviour \footnote{The linear behaviour at a large (negative) bias is likely due to a large series resistance by the p-type ohmic contact.}, as is expected for a p-n junction.
What appears to be a second line consisting of a smaller set of data points arises from the same measurements, but occasionally the timing of the sampling is shifted by one cycle (i.e. the point where t = 0 is shifted) when triggered, and this results in a set of points shifted up from the main curve.
The bottom panel in Fig.~\ref{fig:PNcurrent} shows the current at $V_{\rm{PN}}^{\rm{L}} = -1.8$~V.
The p-n junction current decays as the induced hole gas density decays.

We produce spatial mappings of the electroluminescence from each p-n junction by scanning the piezoelectric stages in the $x$- and $y$-directions.
By varying the applied p-n junction bias, we can operate each p-n junction separately or simultaneously.
The electroluminescence spectrum is taken with 2 second exposure time immediately after the holes are induced in the p-type region.
The electroluminescence mappings in Fig. 3(a) were produced by integrating around the emission peak(s) within a wavelength range of 796.88 nm and 828.04 nm.
The gate voltage $V_{\rm{G}}$ was operated in the same manner as described previously.
When $V_{\rm{PN}}^{\rm{L}} = -1.55$~V and $V_{\rm{PN}}^{\rm{R}} = 0$~V, we observe electroluminescence from what we assume is the left-hand junction [Fig. 3(a) (i)]. 
When $V_{\rm{PN}}^{\rm{R}} = -1.65$~V and $V_{\rm{PN}}^{\rm{L}} = 0$~V, we observe electroluminescence from a different position on the device; we assume that this is where the the right-hand junction is located [Fig. 3(a) (ii)].
The white rectangular box on each plot in Fig. 3(a) is included as a reference for the location of the central mesa for the p-type region with a p-n junction at each end.
The distance and angle between the centres of the two electroluminescence spots match what we expect from the location of the p-type region and the angle of the chip mounted onto the sample holder.
For $V_{\rm{PN}}^{\rm{L}} = -1.55$~V and $V_{\rm{PN}}^{\rm{R}} = -1.65$~V, we observe electroluminescence from both junctions simultaneously [Fig. 3(a) (iii)].
Note that the emission from the right-hand junction appears to be partially obscured in Fig. 3(a) (ii) and (iii); this is due to the fact that the piezoelectric stage had reached the end of its scanning range in the $x$-direction. We later repeated the scan after adjusting the positioning of the lens assembly so that the scanning range fully captured the emission from both junctions, and we can confirm that the emission from the right-hand junction indeed appears circular and similar to that of the left-hand junction.
We also noticed that the right-hand junction required a slightly larger bias and the observed electroluminescence is not as bright as that from the left-hand junction.
We note that the size of the circle does not reflect the actual area of emission. The lithographic size of our p-n junction is only 3~$\upmu$m, much smaller than the observed emission circles, which are attributed to the poor focusing ability of the lens assembly. However, this is advantageous in that the light can be collected from a large area, meaning it is easier to locate the p-n junction.
The small dark spot present on the circular emission in Fig. 3(a) (i) and (iii) is likely due to a defect on the surface of the fibre after polishing.

Figure 3(b) shows accumulated electroluminescence spectra from (i) the left-hand junction and (ii) the right-hand junction as the respective bias voltages were varied (i) from $-1.5$~V to $-1.8$~V and (ii) from $-1.6$~V to $-1.9$~V. The position of the stage was set to (i) $x = 4250$~$\upmu$m, $y = 3900$~$\upmu$m and (ii) $x = 4980$~$\upmu$m, $y = 3360$~$\upmu$m.
The spectra were repeatedly taken 100 times, each using an exposure time of 2 s.
The vertical red line on each plot is placed at 812 nm as a guide, indicating that the centre of the emission peak is at around 812 nm.
This is consistent with the emission due to the recombination of neutral excitons observed in similar quantum well structures \cite{hsiao2020single, dai2014high, koteles1988experimental}.
No other emission peaks were detected in the wavelength range scanned (758 nm -- 982 nm), meaning we can be confident that the emission from our device originates from the GaAs quantum well region.
The full-width at half-maximum value of the emission peak is $\sim 8$~nm, which is a factor of two to four larger than those observed in Refs.~\cite{hsiao2020single} and \cite{dai2014high}.
This may be due to the decaying hole density, affecting the exciton energy.

\begin{figure}[!ht]
    \centering
    \includegraphics[scale=0.45]{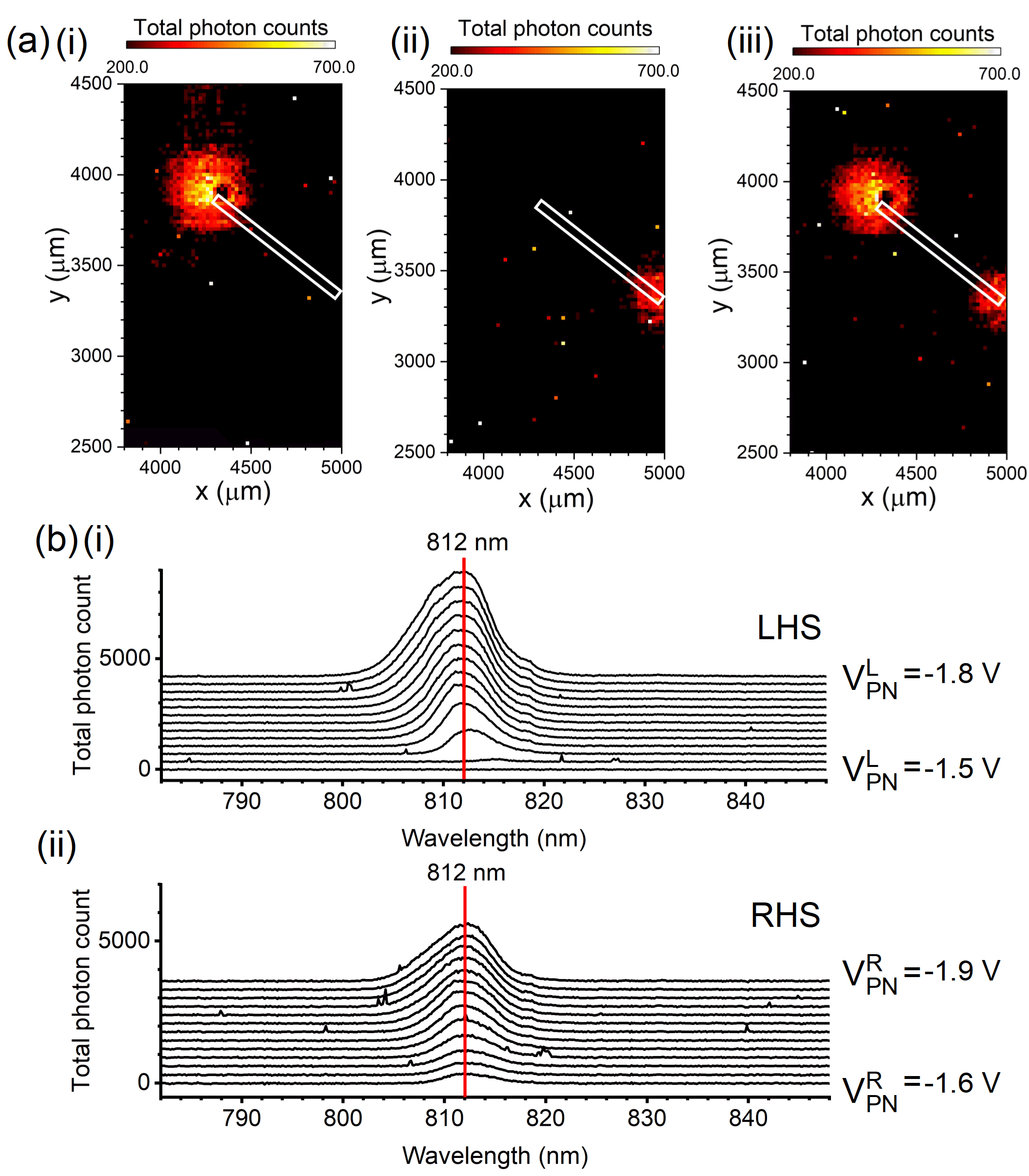}
    \caption{\scriptsize{\textbf{a:} Electroluminescence mappings showing the centres of emission in the x-y plane as the p-n junction biases applied to each junction are varied. \textbf{(i)}: For $V_{\rm{PN}}^{\rm{L}} = -1.55$~V, $V_{\rm{PN}}^{\rm{R}} = 0$~V: emission is assumed to be from the left-hand p-n junction. \textbf{(ii)}: For $V_{\rm{PN}}^{\rm{R}} = -1.65$~V, $V_{\rm{PN}}^{\rm{L}} = 0$~V: emission is assumed to be from the right-hand p-n junction. \textbf{(iii)}: When $V_{\rm{PN}}^{\rm{L}} = -1.55$~V and $V_{\rm{PN}}^{\rm{R}} = -1.65$~V, emission can be seen from both p-n junctions simultaneously. The white bar placed between the centres of emission in each plot in \textbf{(a)} represents the central mesa for the p-type region.
    \textbf{b:} Electroluminescence spectra (accumulated over 100 repeated cycles) from the \textbf{(i)} the left-hand junction and \textbf{(ii)} the right-hand junction as either $V_{\rm{PN}}^{\rm{L}}$ or $V_{\rm{PN}}^{\rm{R}}$ is varied from $-1.5$~V to $-1.8$~V or from $-1.6$~V to $-1.9$~V, respectively. The curves are offset vertically. For the left-hand junction, the stage was positioned at $x = 4250$~$\upmu$m, $y = 3900$~$\upmu$m; for the right-hand junction, the stage was positioned at $x = 4980$~$\upmu$m, $y = 3360$~$\upmu$m.
    The approximate centre of the emission is at 812 nm, as indicated by the red vertical line on each plot.}
    }
    \label{fig:EL}
\end{figure}
In summary, we have demonstrated controlled electroluminescence from the two p-n junctions on our device. 
The junctions are operated by stepping the applied gate voltage $V_{\rm{G}}$ from 7 V to 0 V, to induce holes in the GaAs quantum well layer of the device. 
This induced hole current can survive for only a short amount of time, roughly 1 s, but this is long enough for us to take electroluminescence measurements. 
The p-n junctions were characterised by varying the applied bias and monitoring the behaviour of the current through the junction with time. 
Electroluminescence was observed from each of the left-hand and right-hand p-n junctions separately, or from both simultaneously, by varying the bias voltages applied to each junction. 
We believe that the light from both the left-hand and right-hand junctions is emitted from the GaAs quantum well layer after analysing the electroluminescence spectra of both junctions. 
Because we start with a standard n-type wafer, it is straightforward to add extra layers of gated devices, such as single-electron pumps, in the n-type region.
The electroluminescence observed here is promising for future work to operate the device using electron pumps to inject single electrons over a p-n junction biased below its threshold so that only hot electrons \cite{Fletcher2013,Kataoka2016,Johnson2018} enters the p-type region.
Such a device could be used for a platform for on-demand generation of single photons for applications in quantum communications and imaging.

This research was supported by the UK Department for Business, Energy, and Industrial Strategy.\\
This project has received funding from the European Union's H2020 research and innovation programme under grant agreement no. 862683 ``UltraFastNano". \\

\newpage

\bibliography{bibliography.bib}

\begin{thebibliography}{23}%
\makeatletter
\providecommand \@ifxundefined [1]{%
 \@ifx{#1\undefined}
}%
\providecommand \@ifnum [1]{%
 \ifnum #1\expandafter \@firstoftwo
 \else \expandafter \@secondoftwo
 \fi
}%
\providecommand \@ifx [1]{%
 \ifx #1\expandafter \@firstoftwo
 \else \expandafter \@secondoftwo
 \fi
}%
\providecommand \natexlab [1]{#1}%
\providecommand \enquote  [1]{``#1''}%
\providecommand \bibnamefont  [1]{#1}%
\providecommand \bibfnamefont [1]{#1}%
\providecommand \citenamefont [1]{#1}%
\providecommand \href@noop [0]{\@secondoftwo}%
\providecommand \href [0]{\begingroup \@sanitize@url \@href}%
\providecommand \@href[1]{\@@startlink{#1}\@@href}%
\providecommand \@@href[1]{\endgroup#1\@@endlink}%
\providecommand \@sanitize@url [0]{\catcode `\\12\catcode `\$12\catcode `\&12\catcode `\#12\catcode `\^12\catcode `\_12\catcode `\%12\relax}%
\providecommand \@@startlink[1]{}%
\providecommand \@@endlink[0]{}%
\providecommand \url  [0]{\begingroup\@sanitize@url \@url }%
\providecommand \@url [1]{\endgroup\@href {#1}{\urlprefix }}%
\providecommand \urlprefix  [0]{URL }%
\providecommand \Eprint [0]{\href }%
\providecommand \doibase [0]{https://doi.org/}%
\providecommand \selectlanguage [0]{\@gobble}%
\providecommand \bibinfo  [0]{\@secondoftwo}%
\providecommand \bibfield  [0]{\@secondoftwo}%
\providecommand \translation [1]{[#1]}%
\providecommand \BibitemOpen [0]{}%
\providecommand \bibitemStop [0]{}%
\providecommand \bibitemNoStop [0]{.\EOS\space}%
\providecommand \EOS [0]{\spacefactor3000\relax}%
\providecommand \BibitemShut  [1]{\csname bibitem#1\endcsname}%
\let\auto@bib@innerbib\@empty
\bibitem [{\citenamefont {Foden}\ \emph {et~al.}(2000)\citenamefont {Foden}, \citenamefont {Talyanskii}, \citenamefont {Milburn}, \citenamefont {Leadbeater},\ and\ \citenamefont {Pepper}}]{foden2000high}%
  \BibitemOpen
  \bibfield  {author} {\bibinfo {author} {\bibfnamefont {C.~L.}\ \bibnamefont {Foden}}, \bibinfo {author} {\bibfnamefont {V.}~\bibnamefont {Talyanskii}}, \bibinfo {author} {\bibfnamefont {G.~J.}\ \bibnamefont {Milburn}}, \bibinfo {author} {\bibfnamefont {M.}~\bibnamefont {Leadbeater}},\ and\ \bibinfo {author} {\bibfnamefont {M.}~\bibnamefont {Pepper}},\ }\href@noop {} {\bibfield  {journal} {\bibinfo  {journal} {Physical Review A}\ }\textbf {\bibinfo {volume} {62}},\ \bibinfo {pages} {011803} (\bibinfo {year} {2000})}\BibitemShut {NoStop}%
\bibitem [{\citenamefont {Hosey}\ \emph {et~al.}(2004)\citenamefont {Hosey}, \citenamefont {Talyanskii}, \citenamefont {Vijendran}, \citenamefont {Jones}, \citenamefont {Ward}, \citenamefont {Unitt}, \citenamefont {Norman},\ and\ \citenamefont {Shields}}]{hosey2004lateral}%
  \BibitemOpen
  \bibfield  {author} {\bibinfo {author} {\bibfnamefont {T.}~\bibnamefont {Hosey}}, \bibinfo {author} {\bibfnamefont {V.}~\bibnamefont {Talyanskii}}, \bibinfo {author} {\bibfnamefont {S.}~\bibnamefont {Vijendran}}, \bibinfo {author} {\bibfnamefont {G.}~\bibnamefont {Jones}}, \bibinfo {author} {\bibfnamefont {M.}~\bibnamefont {Ward}}, \bibinfo {author} {\bibfnamefont {D.}~\bibnamefont {Unitt}}, \bibinfo {author} {\bibfnamefont {C.}~\bibnamefont {Norman}},\ and\ \bibinfo {author} {\bibfnamefont {A.}~\bibnamefont {Shields}},\ }\href@noop {} {\bibfield  {journal} {\bibinfo  {journal} {Applied Physics Letters}\ }\textbf {\bibinfo {volume} {85}},\ \bibinfo {pages} {491} (\bibinfo {year} {2004})}\BibitemShut {NoStop}%
\bibitem [{\citenamefont {Gell}\ \emph {et~al.}(2006)\citenamefont {Gell}, \citenamefont {Atkinson}, \citenamefont {Bremner}, \citenamefont {Sfigakis}, \citenamefont {Kataoka}, \citenamefont {Anderson}, \citenamefont {Jones}, \citenamefont {Barnes}, \citenamefont {Ritchie}, \citenamefont {Ward} \emph {et~al.}}]{gell2006surface}%
  \BibitemOpen
  \bibfield  {author} {\bibinfo {author} {\bibfnamefont {J.}~\bibnamefont {Gell}}, \bibinfo {author} {\bibfnamefont {P.}~\bibnamefont {Atkinson}}, \bibinfo {author} {\bibfnamefont {S.}~\bibnamefont {Bremner}}, \bibinfo {author} {\bibfnamefont {F.}~\bibnamefont {Sfigakis}}, \bibinfo {author} {\bibfnamefont {M.}~\bibnamefont {Kataoka}}, \bibinfo {author} {\bibfnamefont {D.}~\bibnamefont {Anderson}}, \bibinfo {author} {\bibfnamefont {G.}~\bibnamefont {Jones}}, \bibinfo {author} {\bibfnamefont {C.}~\bibnamefont {Barnes}}, \bibinfo {author} {\bibfnamefont {D.}~\bibnamefont {Ritchie}}, \bibinfo {author} {\bibfnamefont {M.}~\bibnamefont {Ward}}, \emph {et~al.},\ }\href@noop {} {\bibfield  {journal} {\bibinfo  {journal} {Applied Physics Letters}\ }\textbf {\bibinfo {volume} {89}},\ \bibinfo {pages} {243505} (\bibinfo {year} {2006})}\BibitemShut {NoStop}%
\bibitem [{\citenamefont {Hsiao}\ \emph {et~al.}(2020)\citenamefont {Hsiao}, \citenamefont {Rubino}, \citenamefont {Chung}, \citenamefont {Son}, \citenamefont {Hou}, \citenamefont {Pedr{\'o}s}, \citenamefont {Nasir}, \citenamefont {{\'E}thier-Majcher}, \citenamefont {Stanley}, \citenamefont {Phillips}, \citenamefont {Mitchell}, \citenamefont {Griffiths}, \citenamefont {Farrer}, \citenamefont {Ritchie},\ and\ \citenamefont {Ford}}]{hsiao2020single}%
  \BibitemOpen
  \bibfield  {author} {\bibinfo {author} {\bibfnamefont {T.-K.}\ \bibnamefont {Hsiao}}, \bibinfo {author} {\bibfnamefont {A.}~\bibnamefont {Rubino}}, \bibinfo {author} {\bibfnamefont {Y.}~\bibnamefont {Chung}}, \bibinfo {author} {\bibfnamefont {S.-K.}\ \bibnamefont {Son}}, \bibinfo {author} {\bibfnamefont {H.}~\bibnamefont {Hou}}, \bibinfo {author} {\bibfnamefont {J.}~\bibnamefont {Pedr{\'o}s}}, \bibinfo {author} {\bibfnamefont {A.}~\bibnamefont {Nasir}}, \bibinfo {author} {\bibfnamefont {G.}~\bibnamefont {{\'E}thier-Majcher}}, \bibinfo {author} {\bibfnamefont {M.~J.}\ \bibnamefont {Stanley}}, \bibinfo {author} {\bibfnamefont {R.~T.}\ \bibnamefont {Phillips}}, \bibinfo {author} {\bibfnamefont {T.~A.}\ \bibnamefont {Mitchell}}, \bibinfo {author} {\bibfnamefont {J.~P.}\ \bibnamefont {Griffiths}}, \bibinfo {author} {\bibfnamefont {I.}~\bibnamefont {Farrer}}, \bibinfo {author} {\bibfnamefont {D.~A.}\ \bibnamefont {Ritchie}},\ and\ \bibinfo {author} {\bibfnamefont {C.~J.~B.}\ \bibnamefont {Ford}},\ }\href@noop {}
  {\bibfield  {journal} {\bibinfo  {journal} {Nature Communications}\ }\textbf {\bibinfo {volume} {11}},\ \bibinfo {pages} {917} (\bibinfo {year} {2020})}\BibitemShut {NoStop}%
\bibitem [{\citenamefont {Vijendran}\ \emph {et~al.}(1999)\citenamefont {Vijendran}, \citenamefont {Sazio}, \citenamefont {Beere}, \citenamefont {Jones}, \citenamefont {Ritchie},\ and\ \citenamefont {Norman}}]{vijendran1999independently}%
  \BibitemOpen
  \bibfield  {author} {\bibinfo {author} {\bibfnamefont {S.}~\bibnamefont {Vijendran}}, \bibinfo {author} {\bibfnamefont {P.}~\bibnamefont {Sazio}}, \bibinfo {author} {\bibfnamefont {H.}~\bibnamefont {Beere}}, \bibinfo {author} {\bibfnamefont {G.}~\bibnamefont {Jones}}, \bibinfo {author} {\bibfnamefont {D.}~\bibnamefont {Ritchie}},\ and\ \bibinfo {author} {\bibfnamefont {C.}~\bibnamefont {Norman}},\ }\href@noop {} {\bibfield  {journal} {\bibinfo  {journal} {Journal of Vacuum Science \& Technology B: Microelectronics and Nanometer Structures Processing, Measurement, and Phenomena}\ }\textbf {\bibinfo {volume} {17}},\ \bibinfo {pages} {3226} (\bibinfo {year} {1999})}\BibitemShut {NoStop}%
\bibitem [{\citenamefont {Vaccaro}\ \emph {et~al.}(1998)\citenamefont {Vaccaro}, \citenamefont {Ohnishi},\ and\ \citenamefont {Fujita}}]{vaccaro1998light}%
  \BibitemOpen
  \bibfield  {author} {\bibinfo {author} {\bibfnamefont {P.~O.}\ \bibnamefont {Vaccaro}}, \bibinfo {author} {\bibfnamefont {H.}~\bibnamefont {Ohnishi}},\ and\ \bibinfo {author} {\bibfnamefont {K.}~\bibnamefont {Fujita}},\ }\href@noop {} {\bibfield  {journal} {\bibinfo  {journal} {Applied Physics Letters}\ }\textbf {\bibinfo {volume} {72}},\ \bibinfo {pages} {818} (\bibinfo {year} {1998})}\BibitemShut {NoStop}%
\bibitem [{\citenamefont {North}\ \emph {et~al.}(1999)\citenamefont {North}, \citenamefont {Burroughes}, \citenamefont {Burke}, \citenamefont {Shields}, \citenamefont {Norman},\ and\ \citenamefont {Pepper}}]{north1999two}%
  \BibitemOpen
  \bibfield  {author} {\bibinfo {author} {\bibfnamefont {A.}~\bibnamefont {North}}, \bibinfo {author} {\bibfnamefont {J.}~\bibnamefont {Burroughes}}, \bibinfo {author} {\bibfnamefont {T.}~\bibnamefont {Burke}}, \bibinfo {author} {\bibfnamefont {A.}~\bibnamefont {Shields}}, \bibinfo {author} {\bibfnamefont {C.~E.}\ \bibnamefont {Norman}},\ and\ \bibinfo {author} {\bibfnamefont {M.}~\bibnamefont {Pepper}},\ }\href@noop {} {\bibfield  {journal} {\bibinfo  {journal} {IEEE Journal of Quantum Electronics}\ }\textbf {\bibinfo {volume} {35}},\ \bibinfo {pages} {352} (\bibinfo {year} {1999})}\BibitemShut {NoStop}%
\bibitem [{\citenamefont {Cecchini}\ \emph {et~al.}(2003)\citenamefont {Cecchini}, \citenamefont {Piazza}, \citenamefont {Beltram}, \citenamefont {Lazzarino}, \citenamefont {Ward}, \citenamefont {Shields}, \citenamefont {Beere},\ and\ \citenamefont {Ritchie}}]{cecchini2003high}%
  \BibitemOpen
  \bibfield  {author} {\bibinfo {author} {\bibfnamefont {M.}~\bibnamefont {Cecchini}}, \bibinfo {author} {\bibfnamefont {V.}~\bibnamefont {Piazza}}, \bibinfo {author} {\bibfnamefont {F.}~\bibnamefont {Beltram}}, \bibinfo {author} {\bibfnamefont {M.}~\bibnamefont {Lazzarino}}, \bibinfo {author} {\bibfnamefont {M.}~\bibnamefont {Ward}}, \bibinfo {author} {\bibfnamefont {A.}~\bibnamefont {Shields}}, \bibinfo {author} {\bibfnamefont {H.}~\bibnamefont {Beere}},\ and\ \bibinfo {author} {\bibfnamefont {D.}~\bibnamefont {Ritchie}},\ }\href@noop {} {\bibfield  {journal} {\bibinfo  {journal} {Applied Physics Letters}\ }\textbf {\bibinfo {volume} {82}},\ \bibinfo {pages} {636} (\bibinfo {year} {2003})}\BibitemShut {NoStop}%
\bibitem [{\citenamefont {Cecchini}\ \emph {et~al.}(2004)\citenamefont {Cecchini}, \citenamefont {De~Simoni}, \citenamefont {Piazza}, \citenamefont {Beltram}, \citenamefont {Beere},\ and\ \citenamefont {Ritchie}}]{cecchini2004surface}%
  \BibitemOpen
  \bibfield  {author} {\bibinfo {author} {\bibfnamefont {M.}~\bibnamefont {Cecchini}}, \bibinfo {author} {\bibfnamefont {G.}~\bibnamefont {De~Simoni}}, \bibinfo {author} {\bibfnamefont {V.}~\bibnamefont {Piazza}}, \bibinfo {author} {\bibfnamefont {F.}~\bibnamefont {Beltram}}, \bibinfo {author} {\bibfnamefont {H.}~\bibnamefont {Beere}},\ and\ \bibinfo {author} {\bibfnamefont {D.}~\bibnamefont {Ritchie}},\ }\href@noop {} {\bibfield  {journal} {\bibinfo  {journal} {Applied physics letters}\ }\textbf {\bibinfo {volume} {85}},\ \bibinfo {pages} {3020} (\bibinfo {year} {2004})}\BibitemShut {NoStop}%
\bibitem [{\citenamefont {Cecchini}\ \emph {et~al.}(2005)\citenamefont {Cecchini}, \citenamefont {Piazza}, \citenamefont {Beltram}, \citenamefont {Gevaux}, \citenamefont {Ward}, \citenamefont {Shields}, \citenamefont {Beere},\ and\ \citenamefont {Ritchie}}]{cecchini2005surface}%
  \BibitemOpen
  \bibfield  {author} {\bibinfo {author} {\bibfnamefont {M.}~\bibnamefont {Cecchini}}, \bibinfo {author} {\bibfnamefont {V.}~\bibnamefont {Piazza}}, \bibinfo {author} {\bibfnamefont {F.}~\bibnamefont {Beltram}}, \bibinfo {author} {\bibfnamefont {D.}~\bibnamefont {Gevaux}}, \bibinfo {author} {\bibfnamefont {M.}~\bibnamefont {Ward}}, \bibinfo {author} {\bibfnamefont {A.}~\bibnamefont {Shields}}, \bibinfo {author} {\bibfnamefont {H.}~\bibnamefont {Beere}},\ and\ \bibinfo {author} {\bibfnamefont {D.}~\bibnamefont {Ritchie}},\ }\href@noop {} {\bibfield  {journal} {\bibinfo  {journal} {Applied Physics Letters}\ }\textbf {\bibinfo {volume} {86}},\ \bibinfo {pages} {241107} (\bibinfo {year} {2005})}\BibitemShut {NoStop}%
\bibitem [{\citenamefont {Dai}\ \emph {et~al.}(2014)\citenamefont {Dai}, \citenamefont {Lin}, \citenamefont {Lin}, \citenamefont {Lee}, \citenamefont {Zhang}, \citenamefont {Li},\ and\ \citenamefont {Lee}}]{dai2014high}%
  \BibitemOpen
  \bibfield  {author} {\bibinfo {author} {\bibfnamefont {V.-T.}\ \bibnamefont {Dai}}, \bibinfo {author} {\bibfnamefont {S.-D.}\ \bibnamefont {Lin}}, \bibinfo {author} {\bibfnamefont {S.-W.}\ \bibnamefont {Lin}}, \bibinfo {author} {\bibfnamefont {Y.-S.}\ \bibnamefont {Lee}}, \bibinfo {author} {\bibfnamefont {Y.-J.}\ \bibnamefont {Zhang}}, \bibinfo {author} {\bibfnamefont {L.-C.}\ \bibnamefont {Li}},\ and\ \bibinfo {author} {\bibfnamefont {C.-P.}\ \bibnamefont {Lee}},\ }\href@noop {} {\bibfield  {journal} {\bibinfo  {journal} {Optics Express}\ }\textbf {\bibinfo {volume} {22}},\ \bibinfo {pages} {3811} (\bibinfo {year} {2014})}\BibitemShut {NoStop}%
\bibitem [{\citenamefont {Dai}\ \emph {et~al.}(2012)\citenamefont {Dai}, \citenamefont {Lin}, \citenamefont {Lin}, \citenamefont {Wu}, \citenamefont {Li},\ and\ \citenamefont {Lee}}]{dai2012lateral}%
  \BibitemOpen
  \bibfield  {author} {\bibinfo {author} {\bibfnamefont {V.-T.}\ \bibnamefont {Dai}}, \bibinfo {author} {\bibfnamefont {S.-D.}\ \bibnamefont {Lin}}, \bibinfo {author} {\bibfnamefont {S.-W.}\ \bibnamefont {Lin}}, \bibinfo {author} {\bibfnamefont {J.-Y.}\ \bibnamefont {Wu}}, \bibinfo {author} {\bibfnamefont {L.-C.}\ \bibnamefont {Li}},\ and\ \bibinfo {author} {\bibfnamefont {C.-P.}\ \bibnamefont {Lee}},\ }\href@noop {} {\bibfield  {journal} {\bibinfo  {journal} {Japanese Journal of Applied Physics}\ }\textbf {\bibinfo {volume} {52}},\ \bibinfo {pages} {014001} (\bibinfo {year} {2012})}\BibitemShut {NoStop}%
\bibitem [{\citenamefont {Helgers}\ \emph {et~al.}(2019)\citenamefont {Helgers}, \citenamefont {Sanada}, \citenamefont {Kunihashi}, \citenamefont {Rubino}, \citenamefont {Ford}, \citenamefont {Biermann},\ and\ \citenamefont {Santos}}]{helgers2019sidewall}%
  \BibitemOpen
  \bibfield  {author} {\bibinfo {author} {\bibfnamefont {P.~L.}\ \bibnamefont {Helgers}}, \bibinfo {author} {\bibfnamefont {H.}~\bibnamefont {Sanada}}, \bibinfo {author} {\bibfnamefont {Y.}~\bibnamefont {Kunihashi}}, \bibinfo {author} {\bibfnamefont {A.}~\bibnamefont {Rubino}}, \bibinfo {author} {\bibfnamefont {C.~J.}\ \bibnamefont {Ford}}, \bibinfo {author} {\bibfnamefont {K.}~\bibnamefont {Biermann}},\ and\ \bibinfo {author} {\bibfnamefont {P.~V.}\ \bibnamefont {Santos}},\ }\href@noop {} {\bibfield  {journal} {\bibinfo  {journal} {Physical Review Applied}\ }\textbf {\bibinfo {volume} {11}},\ \bibinfo {pages} {064017} (\bibinfo {year} {2019})}\BibitemShut {NoStop}%
\bibitem [{\citenamefont {Frisenda}\ \emph {et~al.}(2018)\citenamefont {Frisenda}, \citenamefont {Molina-Mendoza}, \citenamefont {Mueller}, \citenamefont {Castellanos-Gomez},\ and\ \citenamefont {Van Der~Zant}}]{frisenda2018atomically}%
  \BibitemOpen
  \bibfield  {author} {\bibinfo {author} {\bibfnamefont {R.}~\bibnamefont {Frisenda}}, \bibinfo {author} {\bibfnamefont {A.~J.}\ \bibnamefont {Molina-Mendoza}}, \bibinfo {author} {\bibfnamefont {T.}~\bibnamefont {Mueller}}, \bibinfo {author} {\bibfnamefont {A.}~\bibnamefont {Castellanos-Gomez}},\ and\ \bibinfo {author} {\bibfnamefont {H.~S.}\ \bibnamefont {Van Der~Zant}},\ }\href@noop {} {\bibfield  {journal} {\bibinfo  {journal} {Chemical Society Reviews}\ }\textbf {\bibinfo {volume} {47}},\ \bibinfo {pages} {3339} (\bibinfo {year} {2018})}\BibitemShut {NoStop}%
\bibitem [{\citenamefont {Chung}\ \emph {et~al.}(2019)\citenamefont {Chung}, \citenamefont {Hou}, \citenamefont {Son}, \citenamefont {Hsiao}, \citenamefont {Nasir}, \citenamefont {Rubino}, \citenamefont {Griffiths}, \citenamefont {Farrer}, \citenamefont {Ritchie},\ and\ \citenamefont {Ford}}]{chung2019quantized}%
  \BibitemOpen
  \bibfield  {author} {\bibinfo {author} {\bibfnamefont {Y.}~\bibnamefont {Chung}}, \bibinfo {author} {\bibfnamefont {H.}~\bibnamefont {Hou}}, \bibinfo {author} {\bibfnamefont {S.-K.}\ \bibnamefont {Son}}, \bibinfo {author} {\bibfnamefont {T.-K.}\ \bibnamefont {Hsiao}}, \bibinfo {author} {\bibfnamefont {A.}~\bibnamefont {Nasir}}, \bibinfo {author} {\bibfnamefont {A.}~\bibnamefont {Rubino}}, \bibinfo {author} {\bibfnamefont {J.~P.}\ \bibnamefont {Griffiths}}, \bibinfo {author} {\bibfnamefont {I.}~\bibnamefont {Farrer}}, \bibinfo {author} {\bibfnamefont {D.~A.}\ \bibnamefont {Ritchie}},\ and\ \bibinfo {author} {\bibfnamefont {C.~J.}\ \bibnamefont {Ford}},\ }\href@noop {} {\bibfield  {journal} {\bibinfo  {journal} {Physical Review B}\ }\textbf {\bibinfo {volume} {100}},\ \bibinfo {pages} {245401} (\bibinfo {year} {2019})}\BibitemShut {NoStop}%
\bibitem [{\citenamefont {Mondal}\ \emph {et~al.}(2014)\citenamefont {Mondal}, \citenamefont {Gardner}, \citenamefont {Watson}, \citenamefont {Fallahi}, \citenamefont {Yacoby},\ and\ \citenamefont {Manfra}}]{Mondal2014}%
  \BibitemOpen
  \bibfield  {author} {\bibinfo {author} {\bibfnamefont {S.}~\bibnamefont {Mondal}}, \bibinfo {author} {\bibfnamefont {G.~C.}\ \bibnamefont {Gardner}}, \bibinfo {author} {\bibfnamefont {J.~D.}\ \bibnamefont {Watson}}, \bibinfo {author} {\bibfnamefont {S.}~\bibnamefont {Fallahi}}, \bibinfo {author} {\bibfnamefont {A.}~\bibnamefont {Yacoby}},\ and\ \bibinfo {author} {\bibfnamefont {M.~J.}\ \bibnamefont {Manfra}},\ }\href {https://doi.org/https://doi.org/10.1016/j.ssc.2014.08.011} {\bibfield  {journal} {\bibinfo  {journal} {Solid State Communications}\ }\textbf {\bibinfo {volume} {197}},\ \bibinfo {pages} {20} (\bibinfo {year} {2014})}\BibitemShut {NoStop}%
\bibitem [{\citenamefont {Buonacorsi}\ \emph {et~al.}(2021)\citenamefont {Buonacorsi}, \citenamefont {Sfigakis}, \citenamefont {Shetty}, \citenamefont {Tam}, \citenamefont {Kim}, \citenamefont {Harrigan}, \citenamefont {Hohls}, \citenamefont {Reimer}, \citenamefont {Wasilewski},\ and\ \citenamefont {Baugh}}]{buonacorsi2021non}%
  \BibitemOpen
  \bibfield  {author} {\bibinfo {author} {\bibfnamefont {B.}~\bibnamefont {Buonacorsi}}, \bibinfo {author} {\bibfnamefont {F.}~\bibnamefont {Sfigakis}}, \bibinfo {author} {\bibfnamefont {A.}~\bibnamefont {Shetty}}, \bibinfo {author} {\bibfnamefont {M.}~\bibnamefont {Tam}}, \bibinfo {author} {\bibfnamefont {H.}~\bibnamefont {Kim}}, \bibinfo {author} {\bibfnamefont {S.}~\bibnamefont {Harrigan}}, \bibinfo {author} {\bibfnamefont {F.}~\bibnamefont {Hohls}}, \bibinfo {author} {\bibfnamefont {M.}~\bibnamefont {Reimer}}, \bibinfo {author} {\bibfnamefont {Z.}~\bibnamefont {Wasilewski}},\ and\ \bibinfo {author} {\bibfnamefont {J.}~\bibnamefont {Baugh}},\ }\href@noop {} {\bibfield  {journal} {\bibinfo  {journal} {Applied Physics Letters}\ }\textbf {\bibinfo {volume} {119}},\ \bibinfo {pages} {114001} (\bibinfo {year} {2021})}\BibitemShut {NoStop}%
\bibitem [{\citenamefont {Blumenthal}\ \emph {et~al.}(2007)\citenamefont {Blumenthal}, \citenamefont {Kaestner}, \citenamefont {Li}, \citenamefont {Giblin}, \citenamefont {Janssen}, \citenamefont {Pepper}, \citenamefont {Anderson}, \citenamefont {Jones},\ and\ \citenamefont {Ritchie}}]{Blumenthal2007}%
  \BibitemOpen
  \bibfield  {author} {\bibinfo {author} {\bibfnamefont {M.~D.}\ \bibnamefont {Blumenthal}}, \bibinfo {author} {\bibfnamefont {B.}~\bibnamefont {Kaestner}}, \bibinfo {author} {\bibfnamefont {L.}~\bibnamefont {Li}}, \bibinfo {author} {\bibfnamefont {S.}~\bibnamefont {Giblin}}, \bibinfo {author} {\bibfnamefont {T.~J. B.~M.}\ \bibnamefont {Janssen}}, \bibinfo {author} {\bibfnamefont {M.}~\bibnamefont {Pepper}}, \bibinfo {author} {\bibfnamefont {D.}~\bibnamefont {Anderson}}, \bibinfo {author} {\bibfnamefont {G.}~\bibnamefont {Jones}},\ and\ \bibinfo {author} {\bibfnamefont {D.~A.}\ \bibnamefont {Ritchie}},\ }\href {https://doi.org/10.1038/nphys582} {\bibfield  {journal} {\bibinfo  {journal} {Nature Physics}\ }\textbf {\bibinfo {volume} {3}},\ \bibinfo {pages} {343} (\bibinfo {year} {2007})}\BibitemShut {NoStop}%
\bibitem [{\citenamefont {Fletcher}\ \emph {et~al.}(2013)\citenamefont {Fletcher}, \citenamefont {See}, \citenamefont {Howe}, \citenamefont {Pepper}, \citenamefont {Giblin}, \citenamefont {Griffiths}, \citenamefont {Jones}, \citenamefont {Farrer}, \citenamefont {Ritchie}, \citenamefont {Janssen} \emph {et~al.}}]{Fletcher2013}%
  \BibitemOpen
  \bibfield  {author} {\bibinfo {author} {\bibfnamefont {J.~D.}\ \bibnamefont {Fletcher}}, \bibinfo {author} {\bibfnamefont {P.}~\bibnamefont {See}}, \bibinfo {author} {\bibfnamefont {H.}~\bibnamefont {Howe}}, \bibinfo {author} {\bibfnamefont {M.}~\bibnamefont {Pepper}}, \bibinfo {author} {\bibfnamefont {S.~P.}\ \bibnamefont {Giblin}}, \bibinfo {author} {\bibfnamefont {J.~P.}\ \bibnamefont {Griffiths}}, \bibinfo {author} {\bibfnamefont {G.~A.~C.}\ \bibnamefont {Jones}}, \bibinfo {author} {\bibfnamefont {I.}~\bibnamefont {Farrer}}, \bibinfo {author} {\bibfnamefont {D.~A.}\ \bibnamefont {Ritchie}}, \bibinfo {author} {\bibfnamefont {T.~J. B.~M.}\ \bibnamefont {Janssen}}, \emph {et~al.},\ }\href@noop {} {\bibfield  {journal} {\bibinfo  {journal} {Physical Review Letters}\ }\textbf {\bibinfo {volume} {111}},\ \bibinfo {pages} {216807} (\bibinfo {year} {2013})}\BibitemShut {NoStop}%
\bibitem [{Note1()}]{Note1}%
  \BibitemOpen
  \bibinfo {note} {The linear behaviour at a large (negative) bias is likely due to a large series resistance by the p-type ohmic contact.}\BibitemShut {Stop}%
\bibitem [{\citenamefont {Koteles}\ and\ \citenamefont {Chi}(1988)}]{koteles1988experimental}%
  \BibitemOpen
  \bibfield  {author} {\bibinfo {author} {\bibfnamefont {E.~S.}\ \bibnamefont {Koteles}}\ and\ \bibinfo {author} {\bibfnamefont {J.}~\bibnamefont {Chi}},\ }\href@noop {} {\bibfield  {journal} {\bibinfo  {journal} {Physical Review B}\ }\textbf {\bibinfo {volume} {37}},\ \bibinfo {pages} {6332} (\bibinfo {year} {1988})}\BibitemShut {NoStop}%
\bibitem [{\citenamefont {Kataoka}\ \emph {et~al.}(2016)\citenamefont {Kataoka}, \citenamefont {Johnson}, \citenamefont {Emary}, \citenamefont {See}, \citenamefont {Griffiths}, \citenamefont {Jones}, \citenamefont {Farrer}, \citenamefont {Ritchie}, \citenamefont {Pepper},\ and\ \citenamefont {Janssen}}]{Kataoka2016}%
  \BibitemOpen
  \bibfield  {author} {\bibinfo {author} {\bibfnamefont {M.}~\bibnamefont {Kataoka}}, \bibinfo {author} {\bibfnamefont {N.}~\bibnamefont {Johnson}}, \bibinfo {author} {\bibfnamefont {C.}~\bibnamefont {Emary}}, \bibinfo {author} {\bibfnamefont {P.}~\bibnamefont {See}}, \bibinfo {author} {\bibfnamefont {J.~P.}\ \bibnamefont {Griffiths}}, \bibinfo {author} {\bibfnamefont {G.~A.~C.}\ \bibnamefont {Jones}}, \bibinfo {author} {\bibfnamefont {I.}~\bibnamefont {Farrer}}, \bibinfo {author} {\bibfnamefont {D.~A.}\ \bibnamefont {Ritchie}}, \bibinfo {author} {\bibfnamefont {M.}~\bibnamefont {Pepper}},\ and\ \bibinfo {author} {\bibfnamefont {T.~J. B.~M.}\ \bibnamefont {Janssen}},\ }\href {https://doi.org/https://doi.org/10.1103/physrevlett.116.126803} {\bibfield  {journal} {\bibinfo  {journal} {Physical Review Letters}\ }\textbf {\bibinfo {volume} {116}},\ \bibinfo {pages} {126803} (\bibinfo {year} {2016})}\BibitemShut {NoStop}%
\bibitem [{\citenamefont {Johnson}\ \emph {et~al.}(2018)\citenamefont {Johnson}, \citenamefont {Emary}, \citenamefont {Ryu}, \citenamefont {Sim}, \citenamefont {See}, \citenamefont {Fletcher}, \citenamefont {Griffiths}, \citenamefont {Jones}, \citenamefont {Farrer}, \citenamefont {Ritchie}, \citenamefont {Pepper}, \citenamefont {Janssen},\ and\ \citenamefont {Kataoka}}]{Johnson2018}%
  \BibitemOpen
  \bibfield  {author} {\bibinfo {author} {\bibfnamefont {N.}~\bibnamefont {Johnson}}, \bibinfo {author} {\bibfnamefont {C.}~\bibnamefont {Emary}}, \bibinfo {author} {\bibfnamefont {S.}~\bibnamefont {Ryu}}, \bibinfo {author} {\bibfnamefont {H.-S.}\ \bibnamefont {Sim}}, \bibinfo {author} {\bibfnamefont {P.}~\bibnamefont {See}}, \bibinfo {author} {\bibfnamefont {J.~D.}\ \bibnamefont {Fletcher}}, \bibinfo {author} {\bibfnamefont {J.~P.}\ \bibnamefont {Griffiths}}, \bibinfo {author} {\bibfnamefont {G.~A.~C.}\ \bibnamefont {Jones}}, \bibinfo {author} {\bibfnamefont {I.}~\bibnamefont {Farrer}}, \bibinfo {author} {\bibfnamefont {D.~A.}\ \bibnamefont {Ritchie}}, \bibinfo {author} {\bibfnamefont {M.}~\bibnamefont {Pepper}}, \bibinfo {author} {\bibfnamefont {T.~J. B.~M.}\ \bibnamefont {Janssen}},\ and\ \bibinfo {author} {\bibfnamefont {M.}~\bibnamefont {Kataoka}},\ }\href {https://doi.org/https://doi.org/10.1103/PhysRevLett.121.137703} {\bibfield  {journal} {\bibinfo  {journal} {{Physical Review Letters}}\ }\textbf
  {\bibinfo {volume} {{121}}},\ \bibinfo {pages} {{137703}} (\bibinfo {year} {{2018}})}\BibitemShut {NoStop}%
\end{thebibliography}%

\end{document}